# Optical signatures of -1/3 fractional quantum anomalous Hall state in twisted MoTe$_2$

Haiyang Pan[1,2,†], Shunshun Yang[1,3,†], Yuzhu Wang[2,†], Xiangbin Cai[1,2,†], Wei Wang[4], Yan Zhao[1,2], Kenji Watanabe[5], Takashi Taniguchi[6], Linlong Zhang[3], Youwen Liu[3], Bo Yang[2*], Weibo Gao[1,2,7,8*]

[1]*School of Electrical & Electronic Engineering, Nanyang Technological University, Singapore 639798, Singapore*

[2]*Division of Physics and Applied Physics, School of Physical and Mathematical Sciences, Nanyang Technological University, Singapore 637371, Singapore*

[3]*College of Physics, Nanjing University of Aeronautics and Astronautics, Key Laboratory of Aerospace Information Materials and Physics (NUAA), Nanjing 211106, China.*

[4]*Key Laboratory of Flexible Electronics (KLoFE) & Institute of Advanced Materials (IAM), School of Flexible Electronics (Future Technologies), Nanjing Tech University, Nanjing 211816, China*

[5]*Research Center for Functional Materials, National Institute for Materials Science, Tsukuba, Japan*

[6]*International Center for Materials Nanoarchitectonics, National Institute for Materials Science, Tsukuba, Japan*

[7]*Centre for Quantum Technologies, Nanyang Technological University, Singapore, Singapore*

[8]*Quantum Science and Engineering Centre (QSec), Nanyang Technological University, Singapore, Singapore*

[†]*These authors contributed equally: Haiyang Pan, Shunshun Yang, Yuzhu Wang, Xiangbin Cai*

[*]*Correspondence to B.Y. (yang.bo@ntu.edu.sg) and W.G. (wbgao@ntu.edu.sg).*

**Abstract**

The discovery of fractional charge excitations in new platforms offers crucial insights into strongly correlated quantum phases. While a range of fractional quantum anomalous Hall (FQAH) states have recently been observed in two-dimensional twisted moiré systems, the theoretically anticipated filling factor $v = -1/3$ FQAH state has remained elusive, with debates centering on its nature of charge density wave or a topological Chern insulator. Here, we report the optical detection of a $v = -1/3$ FQAH state in twisted MoTe$_2$ bilayers. Using photoluminescence (PL) and reflective magnetic circular dichroism (RMCD) techniques, we identify ferromagnetic states at filling factors $v = -1, -2/3$, and $-1/3$, all tunable by a vertical electric field. The corresponding Curie temperatures are approximately 11 K, 3.5 K, and 2.4 K, respectively. The $-1/3$ state emerges over a narrower electric field range and a lower temperature compared to the integer and other fractional states, indicating its fragile nature that may lead to its absence in previous reports. Notably, the PL spectra at $v = -1/3$ disperse as the out-of-plane magnetic field increases, consistent with a nontrivial topological origin. Theoretical calculations based on the exact diagonalization method further support the interpretation of this topologically non-trivial state.

# Introduction

The discovery of fractional charge excitations in two-dimensional electron systems marked a milestone in our understanding of quantum matter. These excitations, arising from strong electronic correlations, were first revealed in the fractional quantum Hall (FQH) effect when a two-dimensional electron gas (2DEG) is subjected to strong perpendicular magnetic fields at cryogenic temperatures[1]. One of the foundational theoretical paradigms for FQH physics was established by Laughlin, whose seminal wavefunction description captured the essential physics of the incompressible quantum fluid on a Landau level at filling factor $v_L = 1/(2p+1)$ with $p$ as a positive integer and predicted the emergence of fractionally charged quasiparticles[2]. Subsequent developments by Jain introduced the composite fermion (CF) picture, in which electrons bind an even number of magnetic flux quanta to form new emergent fermions that experience a reduced effective magnetic field[3]. This framework explains the hierarchy of FQH states as integer quantum Hall effects of composite fermions for the characteristic Jain sequence of filling factors $v_L = m/(2m \pm 1)$, where $m$ is an integer. Among these excitations, the filling factor $v_L = 1/3$ state[4] corresponding to $m=1$ is the first observed and the most robust state in the Jain sequence, characterized by a large excitation gap as an exceptionally stable ground state[5,6]. This state has since become a paradigm for understanding interaction-driven quantum phases and continues to serve as a benchmark in the study of strongly correlated systems[7-13].

In recent years, advances in moiré superlattice engineering have enabled the exploration of correlated topological phases in lattice systems without the need for an external magnetic field[14-18]. Notably, fractional quantum anomalous Hall (FQAH) effect, as a defining experimental signature of fractional Chern insulators (FCIs), has been observed in twisted $MoTe_2$ bilayer (t$MoTe_2$)[19-27] and rhombohedral multilayer graphene aligned with hexagonal boron nitride[28-30]. These systems host a series of fractionalized fillings, including moiré filling factor $v = -2/3$, and $-3/5$, which match the Jain sequence found in conventional FQH systems. However, despite this apparent similarity, the foundational $v = -1/3$ state, which serves as the hallmark of Laughlin physics, remains absent across these moiré FCIs platforms. As the tMoTe$_2$ device quality improves[31,32], emerging data from high-quality tMoTe$_2$ devices show resistance anomalies near $v = -1/3$, hinting at incipient correlated behavior, yet clear signatures of spontaneous ferromagnetism or FQAH effect remain absent. Theoretical studies suggest that at this filling, the gap between the two lowest bands is comparable to the interaction scale, enabling substantial band mixing[33]. This band mixing can drive a competition between charge density wave (CDW) order and topological FCI phases, with the interplay between lattice geometry and interactions[33,34]. These findings highlight the importance of further experimental and theoretical investigations to unravel the microscopic mechanisms that govern the $v = -1/3$ state in lattice-based topological systems.

Here, we report the optical signatures of the $v = -1/3$ FQAH state in tMoTe$_2$. Using a photoluminescence (PL)-based sensing technique, we identify a sequence of integer and fractional quantum anomalous Hall states at $v = -1$, $-2/3$, and $-1/3$. Reflective magnetic circular dichroism (RMCD) measurements reveal electric field-tunable ferromagnetism associated with these states. At $v = -1/3$, temperature-dependent RMCD measurements yield a Curie temperature of 2.4 K, lower than that of the $v = -2/3$ state, suggesting a weaker spin exchange strength. Furthermore, the carrier density corresponding to $v = -1/3$ evolves linearly with magnetic field, consistent with a topologically non-trivial origin, in agreement with our theoretical calculations.

**Results and discussion**

We fabricated the rhombohedral-stacked MoTe$_2$ bilayer device with dual gate structures for optical measurements (Fig. S1), enabling the independent control over carrier density and electric field (see Supplemental Material for details). The targeted twist angle was set to 3.5°, within the 3°~ 4° range where ferromagnetic states emerge[19-21,26,27], enabling access to topological moiré flat bands and associated integer quantum anomalous Hall (QAH) and FQAH phases. As reported previously, the moiré potential of the rhombohedral-stacked MoTe$_2$ bilayer gives rise to two degenerate energy minima, forming a honeycomb lattice with the sublattice pseudospin locked to the layer pseudospin [Fig. 1(a)], thereby realizing the Haldane model with strong Coulomb interactions[35-37]. The band structure for a 3.5° twist angle, as calculated, is shown in Fig. 1(b). The first and second valence moiré bands are separated by an interaction gap, with the first flat moiré band carrying a non-zero Chern number of $C = -1$. Depending on the filling of the topological moiré flat band, both integer quantum anomalous Hall (QAH) and FQAH states can be realized. For the filling factor of $v = -1/3$, only one-third of the first moiré valence band is occupied by holes. An intuitive way to understand this state on a Landau level is through CF theory[3], in which one electron is bound to two effective flux quanta, forming a configuration equivalent to a fully filled CF level, as illustrated schematically at the top of Fig. 1(a). Although the mathematical implementation differs from that of Landau levels due to band geometry and lattice effects, the CF picture can be generalized to moiré systems by recognizing that the Berry curvature induces an effective flux within each moiré unit cell, even in the absence of an external magnetic field[38-40].

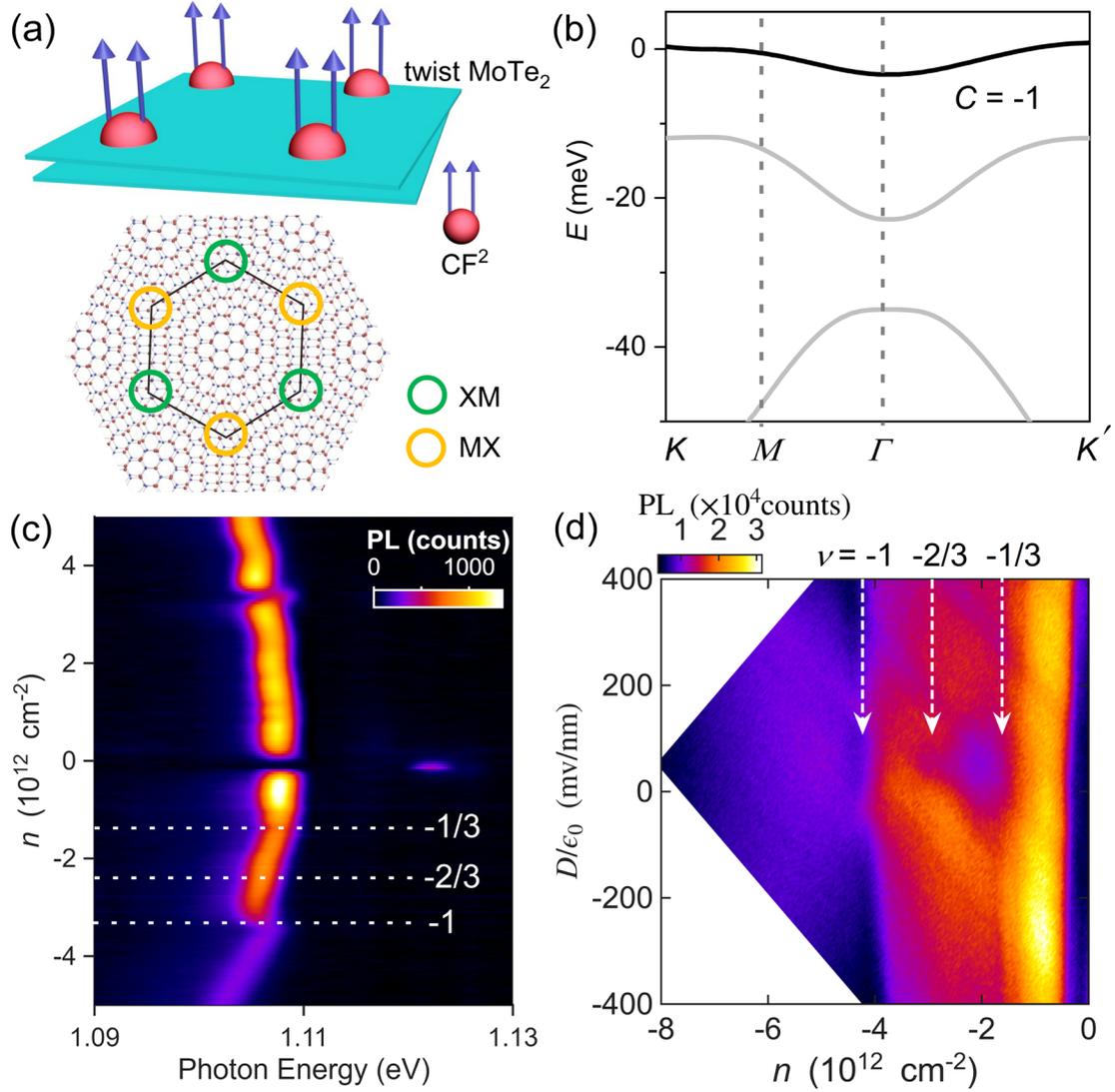

FIG. 1. Optical signatures of $v = -1/3$ state in twisted MoTe$_2$ bilayer. (a) Schematic illustration of the composite fermion (CF) picture for the electronic Laughlin state at 1/3 filling in twisted MoTe$_2$ bilayer (tMoTe$_2$). In this picture, each electron (red ball) binds two effective flux quanta (purple arrows) emerging from band topology, forming a weak-coupling composite fermion, denoted by CF$^2$. A honeycomb moiré superlattice is formed in tMoTe$_2$ composed of two degenerate moiré orbitals localized at XM (green) and MX (yellow) sites in opposite layers. (b) The calculated band structure of tMoTe$_2$ at twist angle of 3.5°. The black line represents the first moiré band, and the gray lines represent the second and third moiré bands. The Chern number of the first moiré miniband is $C = -1$. (c) Photoluminescence (PL) intensity plot as a function of doping carrier density $n$ and photon energy. The white dotted lines indicate the filling states

−1/3, −2/3, and −1. (d) PL intensity versus carrier density and perpendicular electric field $D/\varepsilon_0$ at the hole-doped side. The PL Intensity is integrated over the whole spectrum after a 980-nm long-pass filter. Filling factors $v$ = −1/3, −2/3, and −1 at the hole side are indicated by white dashed arrows.

Figure 1(c) presents the PL intensity as a function of carrier density $n$ and photon energy, measured at a base temperature of 1.5 K unless otherwise specified. As the doping level is tuned from the neutral point to either the hole- or electron-doped sides, the PL intensity exhibits sharp suppression accompanied by a blueshift in the emission peak when a correlated phase forms. The reduction of PL intensity comes from the gap opening at correlated phases, which reduces the free carrier density for the trion emission. The energy-jump feature is attributed to the charge–trion repulsion, which arises when incompressible electronic states form in the moiré superlattice[19,41]. The characteristics enable the designation of different filling factors $v$, defined as the number of charges per moiré unit cell. Prominent features are observed at $v$ = −1/3, −2/3, and −1 on the hole-doped side [white dashed lines, Fig. 1(c)]. In the present study, we focus on the hole-doped regime, where interaction-driven topological states are expected to emerge. The electron-doped side, in contrast, shows no signatures of ferromagnetism and appears topologically trivial. Figure 1(d) displays the PL intensity as a function of carrier density and electric field $D/\varepsilon_0$ on the hole-doped side. From the positions of the integer filling states at $v$ = ±1, we extract a moiré density $n_{moiré}$ = $3.59\times10^{12}$ cm$^{-2}$, corresponding to an effective twist angle of 3.5°.

We performed RMCD measurements to reveal the underlying magnetic property. Figure 2(a) displays the RMCD signal as a function of carrier density $n$ and electric field $D/\varepsilon_0$. Strong RMCD signals emerge on the hole-doped side, consistent with prior observations of ferromagnetic ordering near integer fillings[19,25,41]. Notably, we observe a robust and non-vanishing RMCD signal extending to the fractional filling $v$ = −1/3, indicating ferromagnetism at this fractional state. This magnetic response at $v$ = −1/3 contrasts sharply with earlier reports[19,25,41], where the ferromagnetic signatures were absent at such fractional filling. The ferromagnetic phase exhibits symmetry to the electric field $D/\varepsilon_0$ and emerges at specific carrier densities. To further elucidate the magnetic behavior, we performed RMCD measurements while sweeping the magnetic field up and down at various carrier densities under zero electric field (Fig. S2). The resulting ΔRMCD signal, the difference between RMCD signals when magnetic field is swept up and down, reveals both residual magnetization and the hysteresis loop width, which are plotted as a function of carrier density $n$ and magnetic field in Fig. 2(b). We find enhanced RMCD responses near correlated insulating states at $v = -1, -2/3$, and $-1/3$, evidencing the emergence of an intrinsically spin-polarized correlated phase at these correlated states. The enhancement is strongest at $v = -1$, consistent with previous reports[19,26], while the signal at $v = -2/3$ exceeds that at $v = -1/3$, suggesting a comparatively weaker ferromagnetic order in the latter. This relative weakness may account for the absence of detectable ferromagnetism at $v = -1/3$ in earlier studies[19-21]. Nonetheless, the hysteresis loops shown in Fig. 2(c) exhibit pronounced switching behavior at $v = -1/3$, with a sharp RMCD reversal, providing direct evidence for ferromagnetic ordering at this fractional filling. Additional hysteresis loops at $v = -1/2$ to $-1$ are presented in Figs. 2(d)~2(f), with the signals at $v = -1$ and $-2/3$ exhibiting particularly strong and stable ferromagnetic responses. At $v = -1/2$, the RMCD response does not show a well-defined hysteresis loop. Instead, the signal reverses sharply across zero magnetic field and reaches saturation at higher magnetic fields. This behavior indicates a strongly reduced coercive field, leading to a negligible ΔRMCD, which requires future study.

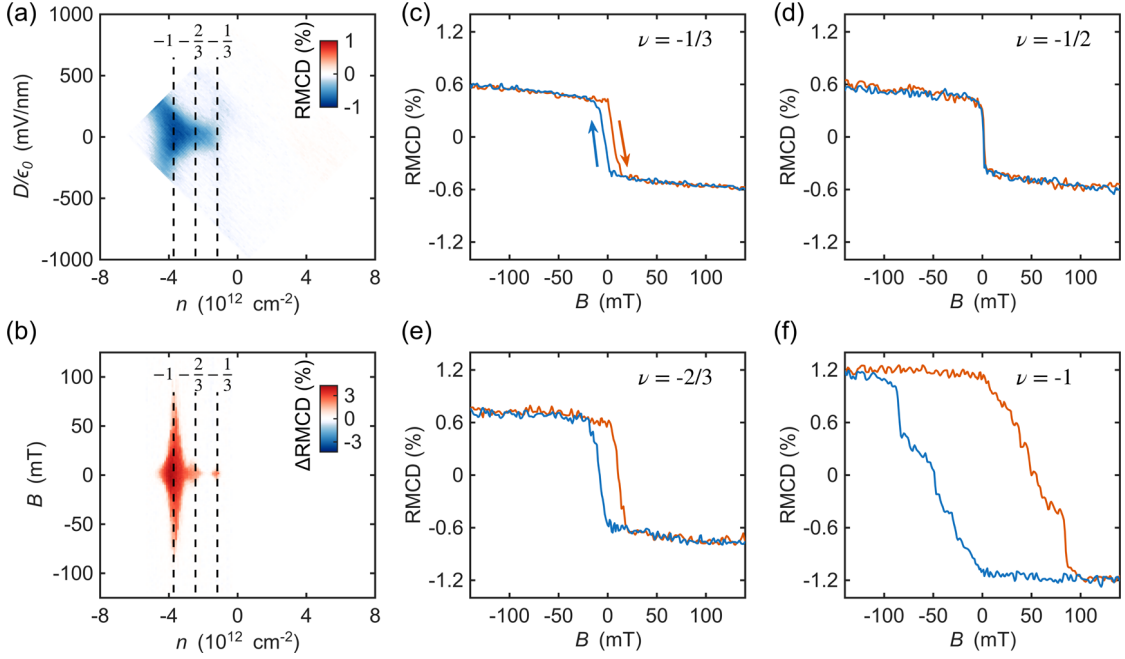

FIG. 2. Ferromagnetism at integer and fractional hole filling states. (a) Reflective magnetic circular dichroism (RMCD) signal measured at 50 mT magnetic field as a function of charge carrier density $n$ and electric field $D/\varepsilon_0$. RMCD measurements were performed at 1.5 K under 35 nW excitation at 1103 nm. (b) Hysteresis component of the RMCD signal as a function of carrier density and magnetic field at zero effective electric field. The differential signal ΔRMCD is defined as the difference in RMCD between the backward and forward magnetic field sweeps. (c)-(f) RMCD signal as a function of magnetic field $B$, swept up (red line) and down (blue line), at filling factors: $v = -1/3$ (c), $v = -1/2$ (d), $v = -2/3$ (e), and $v = -1$ (f). Hysteresis loops indicate spontaneous time-reversal symmetry breaking. All the hysteresis loops were measured at zero effective electric field.

We measured the RMCD signal as a function of electric field $D/\varepsilon_0$ and temperature to investigate the electric and thermal stability of ferromagnetic states at both integer and fractional fillings. Figure 3(a) displays the hysteresis component ΔRMCD at various $D/\varepsilon_0$ for $v = -1$. The ferromagnetic order at $v = -1$ is tunable by the electric field and exhibits a nearly symmetric profile about $D/\varepsilon_0 = 0$. Increasing $D/\varepsilon_0$ suppresses the ferromagnetic signal, which vanishes beyond $|D/\varepsilon_0| \approx 200$ mV/nm. The vertical electric field induces layer polarization, lifting the energy degeneracy between the moiré orbitals localized on the two sublattices, triggering a transition from honeycomb to triangular lattice symmetry and the disappearance of ferromagnetic order. The temperature dependence in Fig. 3(b) shows that the ΔRMCD signal at $v = -1$ persists up to approximately 11 K, consistent with prior reports of the Curie temperature $T_c$[19-21,41].

At fractional filling $v = -2/3$, the electric and thermal dependence of the ΔRMCD signal is shown in Figs. 3(c) and 3(d). Compared to $v = -1$, the hysteresis loop becomes narrower, and the magnetic order is more fragile, disappearing at a smaller electric field strength and a lower temperature. The extracted $T_c$ is about 3.5 K. Crucially, the ferromagnetic response at $v = -1/3$ [Figs. 3(e) and 3(f)] is the weakest among all observed correlated states. The ΔRMCD signal shows a narrow hysteresis area with minimal dependence on the electric field, indicating a relatively fragile magnetic order compared to the ferromagnetic states of $v = -2/3$ and $v = -1$. The ferromagnetic signature vanishes rapidly with increasing temperature, yielding a low Curie temperature of approximately 2.4 K. These observations reveal the fragility of the $v = -1/3$ ferromagnetic state, in sharp contrast to the robustness of the 1/3 FQH state in 2DEG systems, indicating the differences between moiré flat bands and Landau levels. The comparison with the previously observed $v = -2/3$ FCI state shows the generic broken particle-hole symmetry in lattice Chern bands, leading to the reduced spin gap and greater sensitivity to the band mixing effects, and thus the fragility in the $v = -1/3$ state[33,42]. We measured the PL and RMCD signals over a 20 × 20 μm twisted area of this device (Fig. S3), where clear magnetic responses at $v = -1$ and $-2/3$ were observed, indicating good spatial uniformity. RMCD magnetic signals extending to $v = -1/3$ were observed in two distinct regions. Furthermore, ferromagnetic signatures at $v = -1/3$ are reproducible in another tMoTe$_2$ device (Fig. S4). Although the $-1/3$ state is fragile and highly sensitive to factors such as crystal quality and interface conditions, it can still be realized under suitable experimental conditions.

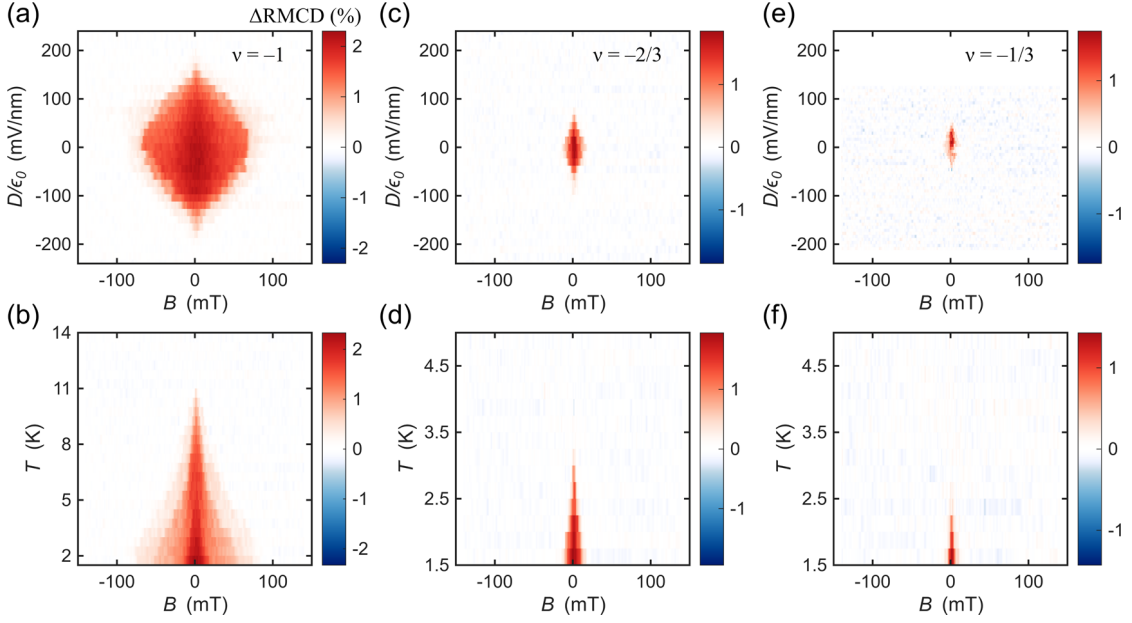

FIG. 3. Electric-field-tunable ferromagnetism and Curie temperature at integer and fractional hole filling states. (a) ΔRMCD, the hysteresis component of the RMCD signal, as a function of electric field $D/\varepsilon_0$ and magnetic field $B$ at integer filling $v = -1$. (b) ΔRMCD signal versus temperature $T$ and $B$ at filling $v = -1$, revealing the temperature dependence of the ferromagnetic state. (c), (d) The same as (a) and (b) but for the fractional filling $v = -2/3$. (e), (f) The same as (a) and (b) but for the fractional filling $v = -1/3$.

We employed a trion-based optical sensing technique to probe the topological character of the ferromagnetic insulating state at fractional filling $v = -1/3$. Figure 4(a) presents the PL spectra as a function of carrier density at varying magnetic fields, measured at zero electric field ($D/\varepsilon_0 = 0$). The emergence of correlated insulating states is characterized by a suppression in trion PL intensity accompanied by a blueshift in its peak energy. Both the $v = -1$ and $v = -1/3$ states exhibit carrier density shifts that scale linearly with magnetic field, with the $v = -1$ slope being steeper. In moiré systems[14,19-21], the Streda formula $C = \frac{h}{e}\frac{\partial n}{\partial B}$ (where $C$ is the Chern number, $h/e$ gives the magnetic flux quantum) provides a way to determine the topological invariant of QAH states. For topologically non-trivial gapped states, the carrier density varies linearly with magnetic field with a quantized slope corresponding to the Chern number, whereas it remains field-independent for trivial states. The $n$ versus $B$ analysis for the $v = -1/3$ and $v = -1$ states (see Supplemental Materials Fig. S5) reveals clear linear evolution. The obtained slopes are qualitatively consistent with the expected Chern sectors, providing evidence that the $v = -1/3$ state is topologically non-trivial.

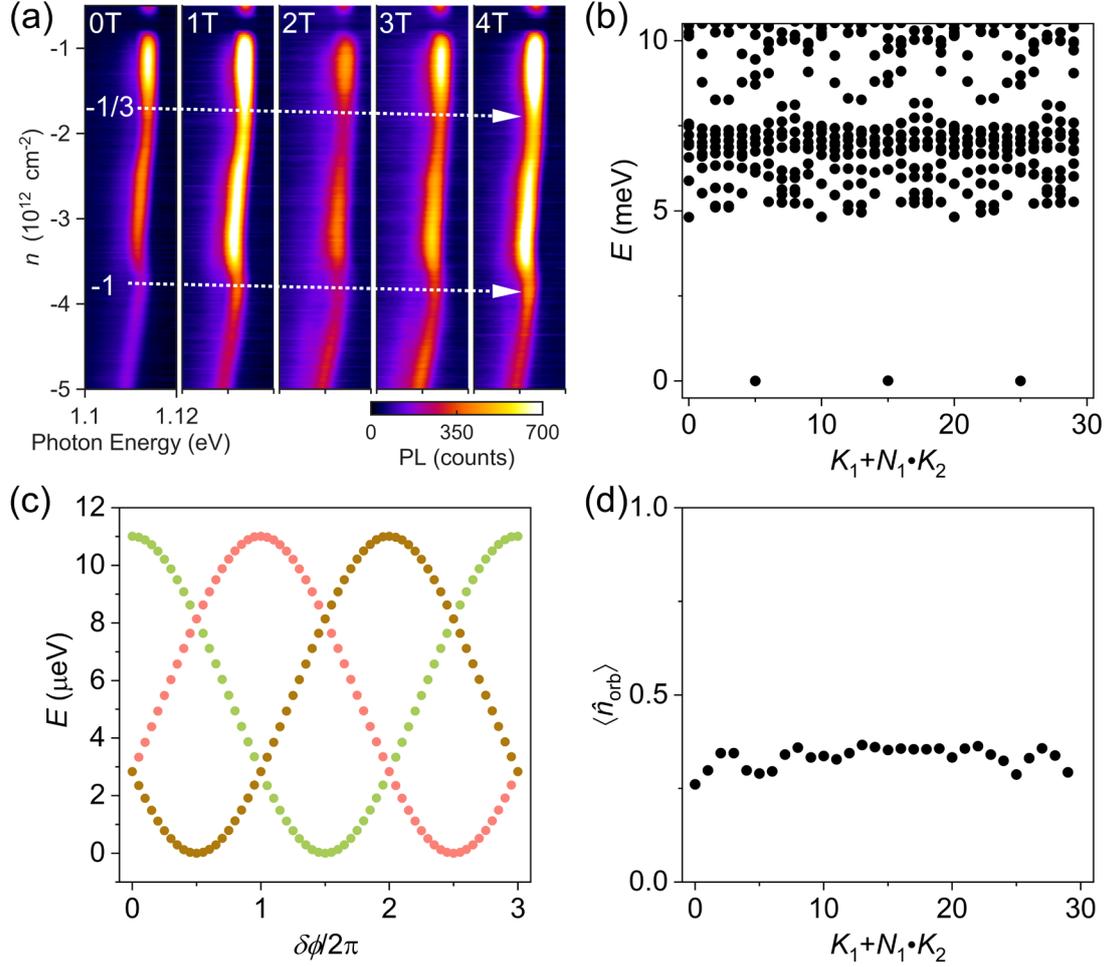

FIG. 4. Nontrivial topology of $v= -1/3$ state. (a) Doping-dependent photoluminescence (PL) spectra at varied magnetic fields (0, 1, 2, 3, 4T from left to right). The white dashed arrows at the top and bottom are the eye guides for $v = -1/3$ and $v = -1$, respectively. (b) Energy spectrum of screened Coulomb interaction on a torus with 30 orbitals in the topmost moiré Chern band shown in Fig. 1(b). Three nearly degenerate ground states are observed, consistent with the expected topological degeneracy of a fractional Chern insulator (FCI) phase. (c) Spectral flow of ground states. The three ground states evolve into one another as a flux $\delta\phi$ is threaded through the torus, while the many-body gap remains open. (d) Orbital occupation $\hat{n}_{\text{orb}}$ distribution of one of the ground states in panel (b). The uniformity supports its identification as an FCI state.

We further carried out exact diagonalization (ED) calculations for the $v = -1/3$ state using realistic material parameters, which strongly support the identification of the observed $v = -1/3$ as an FCI, rather than a trivial CDW. As shown in Fig. 1(b), the single-particle band structure reveals an isolated and flat topmost valence band with Chern number $C = -1$. Such flat Chern bands, characterized by narrow bandwidth and nontrivial topology, are known to favor FCIs by suppressing kinetic energy and enhancing interaction effects[43-48]. Projecting Coulomb interactions into this band at filling $v = -1/3$ yields a many-body spectrum with three nearly degenerate ground states separated by a clear gap (~5 meV) as shown in Fig. 4(b), consistent with the expected topological degeneracy of a Laughlin-like state on a torus[49]. The spectral flow under flux insertion as shown in Fig. 4(c) further confirms the topological nature: the three ground states evolve into each other and return to themselves after three flux quanta are inserted without gap closing[34,42]. Moreover, the orbital density of the ground state is uniform, as demonstrated in Fig. 4(d), ruling out the CDW formation and matching the homogeneity of an FCI ground state[50]. These numerical signatures jointly establish the emergence of an FCI at $v = -1/3$. Details of the ED and theoretical explanations are provided in the Supplementary Materials.

## Conclusions

In conclusion, we have observed optical signatures of a $v = -1/3$ FQAH state in tMoTe$_2$, characterized by a robust yet comparatively weaker ferromagnetic order than its $v = -2/3$ counterpart. Both experimental signatures and theoretical calculations together support an interpretation in terms of a topologically non-trivial fractional state, aligning it with previously reported FCIs. The next key challenge is to detect the -1/3 chiral edge state or the quantized Hall resistance directly through transport measurements, which can provide more direct and compelling evidence of the fractional quantum anomalous Hall state. Continued improvements in device uniformity and contact engineering are expected to enable such experiments. Future studies leveraging advanced spectroscopic techniques and device engineering may reveal even richer fractional phenomena and enable precise control of such exotic states for quantum information applications.

## Acknowledgments

This work is supported by ASTAR (M21K2c0116, M24M8b0004), Singapore National

Research foundation (NRF-CRP22-2019-0004, NRF-CRP30-2023-0003, NRF-CRP31-0001, NRF2023-ITC004-001 and NRF-MSG-2023-0002) and Singapore Ministry of Education Tier 2 Grant (MOE-T2EP50221-0005, MOE-T2EP50222-0018). X.C. acknowledges the support from the NTU Presidential Postdoctoral Fellowship (Grant No. 03INS001828C230). K.W. and T.T. acknowledge support from the JSPS KAKENHI (Grant Numbers 21H05233 and 23H02052) and World Premier International Research Center Initiative (WPI), MEXT, Japan.

## Conflict of Interest

The authors declare no conflict of interest.

## References

[1] H. L. Stormer, Nobel Lecture: The fractional quantum Hall effect, Reviews of Modern Physics **71**, 875 (1999).
[2] R. B. Laughlin, Anomalous Quantum Hall Effect: An Incompressible Quantum Fluid with Fractionally Charged Excitations, Physical Review Letters **50**, 1395 (1983).
[3] J. K. Jain, Composite-fermion approach for the fractional quantum Hall effect, Physical Review Letters **63**, 199 (1989).
[4] D. C. Tsui, H. L. Stormer, and A. C. Gossard, Two-Dimensional Magnetotransport in the Extreme Quantum Limit, Physical Review Letters **48**, 1559 (1982).
[5] D. C. Tsui, H. L. Störmer, J. C. M. Hwang, J. S. Brooks, and M. J. Naughton, Observation of a fractional quantum number, Physical Review B **28**, 2274 (1983).
[6] R. R. Du, H. L. Stormer, D. C. Tsui, L. N. Pfeiffer, and K. W. West, Experimental evidence for new particles in the fractional quantum Hall effect, Physical Review Letters **70**, 2944 (1993).
[7] K. I. Bolotin, F. Ghahari, M. D. Shulman, H. L. Stormer, and P. Kim, Observation of the fractional quantum Hall effect in graphene, Nature **462**, 196 (2009).
[8] D. Kim, S. Jin, T. Taniguchi, K. Watanabe, J. H. Smet, G. Y. Cho, and Y. Kim, Observation of 1/3 fractional quantum Hall physics in balanced large angle twisted bilayer graphene, Nature Communications **16**, 179 (2025).
[9] E. Liu, T. Taniguchi, K. Watanabe, N. M. Gabor, Y.-T. Cui, and C. H. Lui, Excitonic and Valley-Polarization Signatures of Fractional Correlated Electronic Phases in a WSe2/WS2 Moir'e Superlattice, Physical Review Letters **127**, 037402 (2021).
[10] S. Miao *et al.*, Strong interaction between interlayer excitons and correlated electrons in WSe2/WS2 moiré superlattice, Nature Communications **12**, 3608 (2021).
[11] X. Wang *et al.*, Intercell moiré exciton complexes in electron lattices, Nature Materials **22**, 599 (2023).
[12] X. Huang *et al.*, Correlated insulating states at fractional fillings of the WS2/WSe2 moiré lattice, Nature Physics **17**, 715 (2021).
[13] E. C. Regan *et al.*, Mott and generalized Wigner crystal states in WSe2/WS2 moiré

superlattices, Nature **579**, 359 (2020).

[14] M. Serlin, C. L. Tschirhart, H. Polshyn, Y. Zhang, J. Zhu, K. Watanabe, T. Taniguchi, L. Balents, and A. F. Young, Intrinsic quantized anomalous Hall effect in a moiré heterostructure, Science **367**, 900 (2020).

[15] T. Li *et al.*, Quantum anomalous Hall effect from intertwined moiré bands, Nature **600**, 641 (2021).

[16] B. A. Foutty, C. R. Kometter, T. Devakul, A. P. Reddy, K. Watanabe, T. Taniguchi, L. Fu, and B. E. Feldman, Mapping twist-tuned multiband topology in bilayer WSe2, Science **384**, 343 (2024).

[17] K. Kang, B. Shen, Y. Qiu, Y. Zeng, Z. Xia, K. Watanabe, T. Taniguchi, J. Shan, and K. F. Mak, Evidence of the fractional quantum spin Hall effect in moiré MoTe2, Nature **628**, 522 (2024).

[18] K. F. Mak and J. Shan, Semiconductor moiré materials, Nature Nanotechnology **17**, 686 (2022).

[19] J. Cai *et al.*, Signatures of fractional quantum anomalous Hall states in twisted MoTe2, Nature **622**, 63 (2023).

[20] H. Park *et al.*, Observation of fractionally quantized anomalous Hall effect, Nature **622**, 74 (2023).

[21] F. Xu *et al.*, Observation of Integer and Fractional Quantum Anomalous Hall Effects in Twisted Bilayer MoTe2, Physical Review X **13**, 031037 (2023).

[22] Y. Zeng *et al.*, Thermodynamic evidence of fractional Chern insulator in moiré MoTe2, Nature **622**, 69 (2023).

[23] Z. Ji, H. Park, M. E. Barber, C. Hu, K. Watanabe, T. Taniguchi, J.-H. Chu, X. Xu, and Z.-X. Shen, Local probe of bulk and edge states in a fractional Chern insulator, Nature **635**, 578 (2024).

[24] E. Redekop *et al.*, Direct magnetic imaging of fractional Chern insulators in twisted MoTe2, Nature **635**, 584 (2024).

[25] L. An *et al.*, Observation of ferromagnetic phase in the second moiré band of twisted MoTe2, Nature Communications **16**, 5131 (2025).

[26] H. Park *et al.*, Ferromagnetism and topology of the higher flat band in a fractional Chern insulator, Nature Physics **21**, 549 (2025).

[27] F. Xu *et al.*, Interplay between topology and correlations in the second moiré band of twisted bilayer MoTe2, Nature Physics **21**, 542 (2025).

[28] Z. Lu *et al.*, Fractional quantum anomalous Hall effect in multilayer graphene, Nature **626**, 759 (2024).

[29] Z. Lu *et al.*, Extended quantum anomalous Hall states in graphene/hBN moiré superlattices, Nature **637**, 1090 (2025).

[30] J. Xie *et al.*, Tunable fractional Chern insulators in rhombohedral graphene superlattices, Nature Materials (2025).

[31] F. Xu *et al.*, Signatures of unconventional superconductivity near reentrant and fractional quantum anomalous Hall insulators, arXiv:2504.06972 (2025).

[32] H. Park *et al.*, Observation of High-Temperature Dissipationless Fractional Chern Insulator, arXiv:2503.10989 (2025).

[33] Y. He, S. H. Simon, and S. A. Parameswaran, Fractional Chern Insulators and Competing States in a Twisted MoTe Lattice Model, arXiv:2505.06354 (2025).

[34] A. P. Reddy, F. Alsallom, Y. Zhang, T. Devakul, and L. Fu, Fractional quantum anomalous Hall states in twisted bilayer MoTe2 and WSe2, Physical Review B **108**, 085117 (2023).

[35] H. Yu, M. Chen, and W. Yao, Giant magnetic field from moiré induced Berry phase in

homobilayer semiconductors, National Science Review **7**, 12 (2019).

[36] M. H. Naik and M. Jain, Ultraflatbands and Shear Solitons in Moire Patterns of Twisted Bilayer Transition Metal Dichalcogenides, Physical Review Letters **121**, 266401 (2018).

[37] F. Wu, T. Lovorn, E. Tutuc, and A. H. MacDonald, Hubbard Model Physics in Transition Metal Dichalcogenide Moir'e Bands, Physical Review Letters **121**, 026402 (2018).

[38] T. Lu and L. H. Santos, Fractional Chern Insulators in Twisted Bilayer MoTe2: A Composite Fermion Perspective, Physical Review Letters **133**, 186602 (2024).

[39] H. Goldman, A. P. Reddy, N. Paul, and L. Fu, Zero-Field Composite Fermi Liquid in Twisted Semiconductor Bilayers, Physical Review Letters **131**, 136501 (2023).

[40] X. Hu, D. Xiao, and Y. Ran, Hyperdeterminants and composite fermion states in fractional Chern insulators, Physical Review B **109**, 245125 (2024).

[41] E. Anderson, F.-R. Fan, J. Cai, W. Holtzmann, T. Taniguchi, K. Watanabe, D. Xiao, W. Yao, and X. Xu, Programming correlated magnetic states with gate-controlled moiré geometry, Science **381**, 325 (2023).

[42] C. Wang, X.-W. Zhang, X. Liu, Y. He, X. Xu, Y. Ran, T. Cao, and D. Xiao, Fractional Chern Insulator in Twisted Bilayer MoTe2, Physical Review Letters **132**, 036501 (2024).

[43] N. Regnault and B. A. Bernevig, Fractional Chern Insulator, Physical Review X **1**, 021014 (2011).

[44] D. N. Sheng, Z.-C. Gu, K. Sun, and L. Sheng, Fractional quantum Hall effect in the absence of Landau levels, Nature Communications **2**, 389 (2011).

[45] T. Neupert, L. Santos, C. Chamon, and C. Mudry, Fractional Quantum Hall States at Zero Magnetic Field, Physical Review Letters **106**, 236804 (2011).

[46] K. Sun, Z. Gu, H. Katsura, and S. Das Sarma, Nearly Flatbands with Nontrivial Topology, Physical Review Letters **106**, 236803 (2011).

[47] E. Tang, J.-W. Mei, and X.-G. Wen, High-Temperature Fractional Quantum Hall States, Physical Review Letters **106**, 236802 (2011).

[48] S. A. Parameswaran, R. Roy, and S. L. Sondhi, Fractional quantum Hall physics in topological flat bands, Comptes Rendus Physique **14**, 816 (2013).

[49] X. G. Wen and Q. Niu, Ground-state degeneracy of the fractional quantum Hall states in the presence of a random potential and on high-genus Riemann surfaces, Physical Review B **41**, 9377 (1990).

[50] H. Liu, K. Yang, A. Abouelkomsan, Z. Liu, and E. J. Bergholtz, Broken symmetry in ideal Chern bands, Physical Review B **111**, L201105 (2025).